 \documentclass[aps,prb,twocolumn,groupedaddress,showpacs]{revtex4}

\usepackage{graphicx}

\begin{document}

\title{Formation energy in $\sigma$-phase Fe-V alloys}

\author{J. Cieslak}
\email[Corresponding author: ]{cieslak@fis.agh.edu.pl}
\author{J. Tobola}
\author{S. M. Dubiel}
\affiliation{AGH University of Science and Technology,
             Faculty of Physics and Applied Computer Science,
             al. Mickiewicza 30, 30-059 Krakow, Poland}

\date{\today}

\begin{abstract}

Formation energy of the $\sigma$-phase in the Fe-V alloy system, $\Delta E$,
was computed in the full compositional range of its occurrence ($\sim 34 \le x \le \sim 60$)
using the electronic band structure calculations by means of the KKR method.
$\Delta E$-values were found to strongly depend on the Fe concentration,
also its variation with different site occupancies was characteristic of a given lattice site.
Calculated magnetic, $S_{magn}$, and configuration, $S_{conf}$, entropy contributions
were used to determine sublattice occupancies for various compositions and temperatures.
The results agree well with those obtained from neutron diffraction measurements.
\end{abstract}


\maketitle

\section{Introduction }

Sigma ($\sigma$) phase was found in over 40 binary alloys and many other
three- or multi-component alloy systems \cite{Hall66,Joubert08}.  The phase
cannot be formed at the stage of solidification of the solution of alloying
elements.  Instead, it can only be obtained by a high temperature annealing
process (solid state reaction).  From the viewpoint of technological
applications of alloys, the $\sigma$-phase should be avoided as its presence
drastically deteriorates different mechanical properties of materials in which
it precipitates. A better knowledge of conditions of its formation may be helpful 
for fabrications of materials in which the unwanted phase does not precipitate or 
its precipitatioin is significantly retarded. On the other hand, due to its 
complexity (30 atoms per unit cell, five different sublattices, high (12-15) 
coordination numbers, and lack of a stoichiometry) make the  $\sigma$-phase 
a very challending object for theoretical calculations. They are very helpful 
for a proper interpretation of measurements, especially those done by means of 
microscopic methods like M\"ssbauer spectroscopy or nuclear magnetic resonance, 
as well as for a better understanding of the results obtained.
\cite{Cieslak08b,Cieslak10a,Cieslak12a,Cieslak12b}

In the Fe-V system the $\sigma$-phase can be formed by an isothermal annealing
of the {\it bcc}-master alloys in a wide range of concentration ($\sim 34-60$
at.\% V) and temperature ($\sim 400-1230 ^\circ$C) \cite{Hall66}.  Once
formed, it `remains stable at lower temperatures, but it can be dissolved into
the $\alpha$-phase at temperatures above $\sim 1230^\circ$C.  Mechanism of its
formation as well as that of its dissolution is not known with sufficient
precision yet.  Knowledge of the energy of its formation compared to the
energy of formation of the $\alpha$-phase (from which it precipitates) may
help to better understand processes responsible for its formation and
dissolution.  Thus, it may be useful in a fabrication of a new generation of
materials, like stainless steels, having suitable properties for a
construction of new generations of industrially important facilities.

Having a tetragonal unit cell and hosting 30 atoms distributed
over five crystallographically non-equivalent sites (usually termed A through E)
this phase remains still challenging for theoretical calculations. Luckily, increase
of computing capabilities of modern computers makes it possible that
more sophisticated models can be applied to more adequately and precisely
describe the complex structure and properties of the $\sigma$-phase
\cite{Kabliman09,Pavlu10,Kabliman11a,Kabliman11b}.

\section{Computational Details}

The present analysis of the structural stability of the $\sigma$-phase was based on
electronic structure calculations, similar to the methodology applied for
$\sigma$-FeV and $\sigma$-FeCr alloy systems to study hyperfine interactions
\cite{Cieslak08b,Cieslak10a,Cieslak10b,Cieslak12b} as well as the formation energy
for the latter \cite{Cieslak12a}. We have employed the Korringa-Kohn-Rostoker (KKR)
technique \cite{stopa,cpa,Kaprzyk90} to calculate the electronic structure and
the total energy in the framework of the non-relativistic LDA theory and using the
muffin-tin approximation for a crystal potential. More computational details on $\sigma-$phase
can be found in Refs. \onlinecite{Cieslak08b} and \onlinecite{Cieslak10a}.

Since in the real structure of $\sigma$-FeV all five crystallographic sites
are occupied by both types of atoms, a chemical disorder on these
sublattices must be taken into account.  A more sophisticated
approach based on the KKR method with the coherent potential approximation
(CPA), allowing to treat the disorder as random, should be capable of accounting for $2^{30}$
atomic configurations, a target that still remains unreachable.
In these circumstances, an alternative, but a feasible approach consists in
selection of some ordered elementary unit cells with
different configurations of atoms on the five sublattices, which permits to use the
KKR method (instead of KKR-CPA one).
In practice, the electronic structure
calculations were performed for a finite number of the unit cells with a
symmetry reduced to a simple tetragonal one (P1 instead of P4$_2$/mnm),
in which each position (but not the sublattice) was occupied by one type
of atoms (Fe or V).  The atoms were distributed over the five different sites
with the probabilities determined from neutron diffraction measurements
\cite{Cieslak08c} which were next altered in the way described below.
Differences between total energies (per atom), $\Delta E$ (usually
called the formation energy), determined for each of these atomic
configurations, $E_\sigma$, and corresponding values obtained for pure
constituents, i.e. {\it bcc}-Fe and {\it bcc}-V, $E_{\alpha}$, could be
calculated \cite{Turchi04} according to Eq.  (\ref{eqRS}):
\begin{equation}
  \Delta E = E_\sigma-(1-x)E_\alpha^{Fe}-xE_\alpha^V
\label{eqRS}
\end{equation}
where $x$ stands for the concentration of V.  They can be further presented in
form of a diagram as a function of concentration of the alloying elements.  This
approach was already successfully applied to the analysis of the formation energy of
the $\sigma$-phase in the Fe-Cr alloy system \cite{Cieslak12a}.

As mentioned above, the $\sigma$-phase is characterized by a chemical disorder
on all five sublattices, and the concentrations of Fe/V atoms on each of them
were determined experimentally \cite{Cieslak08c}.  Since both Fe and V atoms
are present on all five sublattices, it is useful for further considerations
to introduce shortened notations for samples with different sublattice
occupancies. Multiplicity (denoted by $N_i$) of the five Wyckoff positions, namely
2a, 4f, 8i, 8i' and 8j correspond to A, B, C, D and E site, respectively.
Since the total number of atoms on each of the five sublattices was already
known, thus it was enough to indicate the number of the Fe atoms on each
sublattice explicitly, while the V atoms constitute the balance.  According to
this convention, an illustrative example of $\sigma$-20172 describes a
system comprising 2, 1, 7, and 2 atoms of Fe on the sites A, C, D and
E, respectively, and also 4, 7, 1 and 6 atoms of V atoms on the sites B, C, D and E.

In the present work we analyzed more carefully ordered unit cells
corresponding to the following atomic configurations:  $\sigma$-20172,
$\sigma$-20383 and $\sigma$-21485. The total number of Fe atoms in them is
12, 16 and 20, respectively.  The three configurations have been chosen
intentionally to best approximate experimentally investigated samples viz.
Fe$_{41}$V$_{59}$, Fe$_{52}$V$_{48}$ and Fe$_{66}$V$_{34}$,
respectively \cite{Cieslak12b}.  These three cases correspond to the border
and middle compositions of the $\sigma$-phase existence on the phase diagram of the Fe-V alloy system
\cite{Kubaschewski82}.  Actually, we treated these configurations as the reference
for a further analysis of $\Delta E$.

\begin{figure}
\includegraphics[width=.49\textwidth]{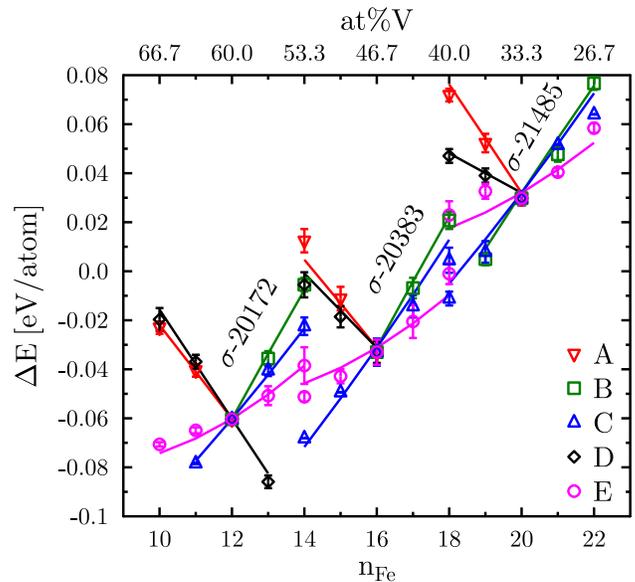}
\caption{(Online color)
Calculated formation energy differences for the Fe-V $\sigma$-phase, $\Delta E$, for the
chosen three basic atomic configurations and its neighborhood, versus the number of Fe
atoms per unit cell, $n_{Fe}$ (V concentration is also indicated).  Solid
lines stand for the best fit to the data - see Appendix.  $\Delta E$-values
determined for occupation changes (with respect to the reference configurations)
on the five lattice sites are indicated by different colors.
}
\label{fig1}
\end{figure}

In the next step we have calculated the $\Delta E$-values, for the unit cells changed
(with respect to the three reference ones) by a replacement of one or two
atoms exclusively on the same site.  Ten ordered approximants of each
configuration of sublattice occupancies with different atomic arrangements
were analyzed, and average $\Delta E$-values, obtained in this approach
for neighborhoods of the three reference configurations, are presented as
a function of Fe atoms number per unit cell, $n_{Fe}$ (Fig.~\ref{fig1}).
Simultaneously, the $\Delta E$-changes caused by the replacements of atoms on each
sublattice are indicated with various colors.

\section{Results and Discussion}

The changes of $\Delta E$ in the vicinity of the three reference
configurations are fairly linear functions of $n_{Fe}$ for the sites A, B, C,
and D, and rather non-linear for the site E.  The slopes of the lines for the sites
A, B and C are characteristic of the site and equal -0.021(1), 0.025(1) and 0.019(1) - in eV/atom -
respectively.  Interestingly, these slopes are more or less the same for all
reference configurations (see, Fig.  \ref{fig1}), while for the site D the value of the slope
depends on the alloy basic configuration around which it is calculated
(changes from -0.022 for $\sigma$-20172 to -0.009 eV/atom for $\sigma$-21485).
$\Delta E$ corresponding to the changes on the site E departs from linearity and
can be approximated using a second-order polynomial of the Fe-concentration on
this site.

The model used for the formation energy calculation allowed to
determine the values of $\Delta E$ only for such cases, in which individual
sites (but not sublattices) are occupied by just one type of atoms.  This
really corresponds to a one selected unit cell, whereas in the neighboring cells
atoms may be arranged in a somewhat different way on the sublattices.  For that
reason, for a more realistic description of crystalline samples it is more
convenient to operate with the average formation energy that corresponds to
the average occupancy of the sublattices by the alloying elements.  As shown
in Fig.  \ref{fig1}, $\Delta E$ have been determined for almost 50 different
configurations of Fe and V atoms, including - to some extend - the chemical disorder
on the sublattices.  The relatively simple character of the $\Delta E$-changes
described above (mostly linear behavior), provides a possibility of determining
a dependence of $\Delta E$ on the Fe-concentrations, $x_i$, $i=$A through E, on each sublattice.
It can be achieved by approximation of $\Delta E$ using a function depending on
the variables $x_i$, on five individual sites, $\Delta E=\Delta
E(x_A,x_B,x_C,x_D,x_E)$.  Taking into account the established character of the
$\Delta E$-changes, there were assumed to be linear (A trough D sites), and quadratic
(E site) dependences on $x_i$ (see Appendix).  Fitting such-defined functions to the
calculated $\Delta E$-values (using the least-square method) yielded numerical
values of polynomial coefficients.  Consequently, it allowed to determine
$\Delta E$ for any atomic configuration for all five sites in the
$\sigma$-FeV structure.

\subsection{Entropy contributions}

Bearing in mind that the aforementioned total energy $E$ and the electronic
structure analysis correspond, in principle, to ground state properties, one can
attempt to discuss crystal stability at finite temperature $T$, applying
commonly used scheme based on a computation of the free energy, $F=E-TS$, where
$S$ stays for an entropy. Considering the process of the $\sigma$-phase formation, one can
discuss $\Delta F = \Delta E - TS$, being the difference between the free
energy of $\sigma$-FeV alloy and free energies of its constituents (namely,
weighted contributions of a pure $bcc$ Fe and V which are already known).

The total entropy that actually should be taken into account in the present case has several
contributions, namely:  magnetic entropy, $S_{mag}$, configuration entropy,
$S_{conf}$, electronic entropy, $S_{el}$, and phonon entropy, $S_{ph}$.  The
first two terms could be computed, and the results of these computations are
shown in Fig.  2, for the magnetic entropy, and in Fig.  3, for the
configuration one.

\begin{figure}
\includegraphics[width=.49\textwidth]{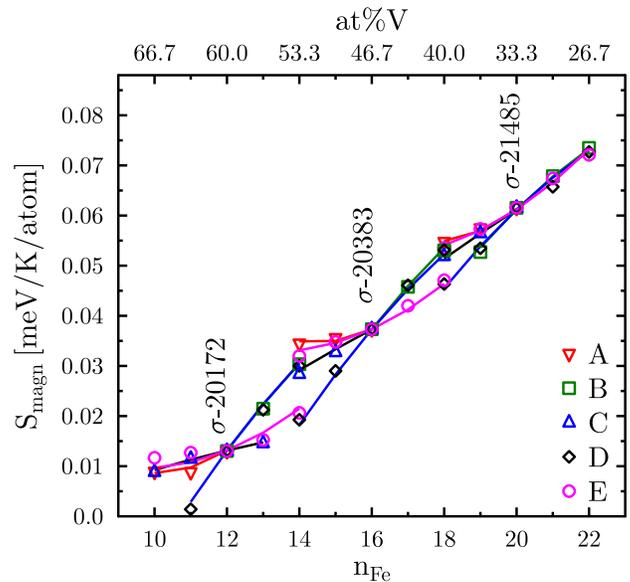}
\caption{(Online color)
The magnetic entropy, $S_{mag}$, calculated versus the number of Fe atoms per unit cell,
$n_{Fe}$, for different atomic configurations (samples) of the $\sigma$-FeV.
The concentration of vanadium is indicated, too. Solid lines stand for the best
fit to the data - see Appendix.  $S_{mag}$-values determined for occupation
changes on the individual lattice site are indicated by different colors.
}
\label{fig2}
\end{figure}

Using the magnetic moments maximum saturation entropy for each atom on
the sublattice, $S_{mag}^{max}=k_B\ln(2J+1)$, can be determined  \cite{Chuang85}.
In this work $S_{mag}$-values were calculated according to the following
Eq.  (\ref{eqSM}):
\begin{equation}
  S_{mag} = k_B\sum (P_i \ln(\mu_{i}^{Fe}+1)+ (1-P_i)\ln(\mu_{i}^{V}+1))
\label{eqSM}
\end{equation}
where $k_B$ is the Boltzmann constant, $P_i$ stands for the occupancy of Fe on
each sublattice, $\mu_i$ is the average magnetic moment of Fe/V atoms
belonging to this sublattice and summation ($i$-index) runs over all five sublattices i.e. from A to E.
$\mu_i$-values for each sublattice and  different $x_i$ configurations
were obtained from the electronic structure KKR calculations.
Fig. 2 gives a clear evidence that for each sample
$S_{mag}$ increases with $n_{Fe}$ for each lattice site.  The increase of
the magnetic entropy remains in line with our previous calculations reporting that
the magnetic moments (also the hyperfine fields) at particular sites increase
with $n_{Fe}$ \cite{Cieslak12b}.  Similarly to the $\Delta E$ expression,
$S_{mag}$ can be represented as a function of $x_i$,
$S_{mag}=S_{mag}(x_A,x_B,x_C,x_D,x_E)$. Although changes in $S_{mag}$ due to changes in the individual
site occupancies are monotonic, however, they deviate slightly from linearity.
To approximate the $S_{mag}$ dependence on $x_i$, we assumed a mathematical form
of the function analogous to that used for $\Delta E$ (see Appendix).  The
numerical values of the polynomial coefficients were determined by
fitting the function to the calculated $S_{mag}$-values.

\begin{figure}
\includegraphics[width=.49\textwidth]{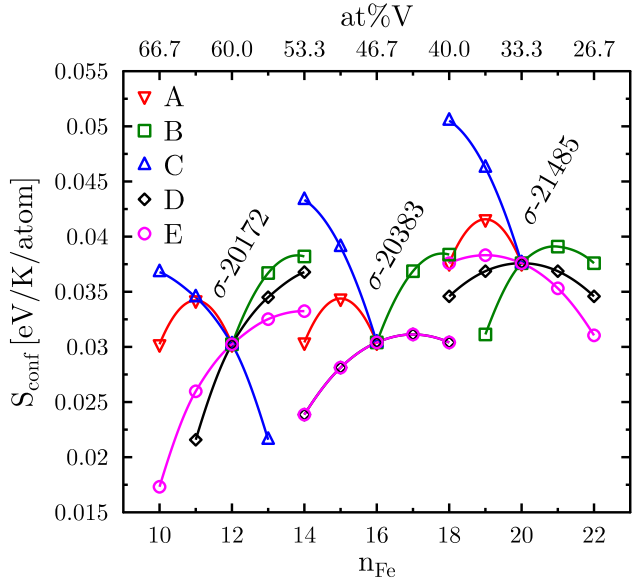}
\caption{(Online color)
The configuration entropy, $S_{conf}$, calculated versus the number of Fe atoms
per unit cell, $n_{Fe}$, for different atomic configurations of the
$\sigma$-FeV.  The concentration of vanadium is indicated, too.
$S_{conf}$-values determined for the occupation changes at each lattice site are
plotted using different colors.
}
\label{fig3}
\end{figure}

The configuration entropy, shown in Fig. 3, was calculated according to Eq.
(\ref{eqSc}):
\begin{equation}
  S_{conf} = -k_B\sum (P_i \ln(P_i)+ (1-P_i)\ln(1-P_i))
\label{eqSc}
\end{equation}
It shows a strong dependence on the site for each of three basic
configurations. The dependence is very similar for the sites A, B and C, while
$S_{conf}$ for the other two sites i.e.  D and E depends on $n_{Fe}$ in a way
characteristic of the reference configuration. Eq.~\ref{eqSc} shows that the
configuration entropy, $S_{conf}$, can be easily calculated for any
concentration of atoms on all sites, $S_{conf}=S_{conf}(x_A,x_B,x_C,x_D,x_E)$.

\begin{figure}
\includegraphics[width=.49\textwidth]{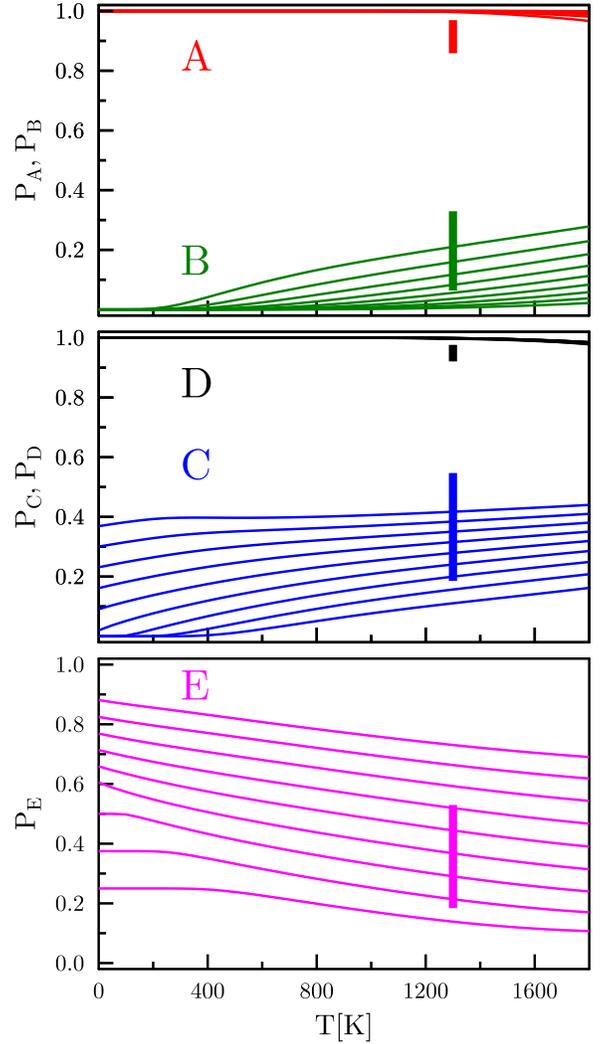}
\caption{(Online color)
The probability of finding Fe-atoms on the five lattice sites, $P_{i}$, versus
temperature, $T$.  The solid lines indicate the calculated values for
different compositions of the $\sigma$-Fe$_{100-x}$V$_x$ with $x$ ranging from
33 to 60 (the curves from top to bottom correspond to alloys with increasing
$x$).  The vertical bars indicate experimentally found values of $P_{i}$ in
the same concentration range.
}
\label{fig4}
\end{figure}

\subsection{Free energy contribution}

The knowledge of $\Delta E$, $S_{mag}$ and $S_{conf}$ for any concentration of
Fe in the sublattices allows to specify a formula describing a change of the free
energy, $\Delta F = \Delta E-T (S_{mag} + S_{conf})$ for any
concentration of the alloying elements, and any temperature, $\Delta F=\Delta
F(x_A,x_B,x_C,x_D,x_E;T)$.  Determination of the lowest value of this function for
the given temperature and concentration, allowed to determine sublattice
occupancies, $P_i$ at the temperature $T$ and Fe-concentration $x_{Fe} = {1\over30}\sum_i
N_i P_i$ (where $N_i$ denotes the occupation number for the sublattice $i$).
The $P_i$-values calculated for some chosen concentrations are shown in Fig. 4
for the temperature ranging between 0 and 1800 K.  At the first glace, one notices a significant site
preference for the occupancies in the investigated phase.  In particular, one can
clearly see that the Fe occupancy of the sites A and D is actually complete and
independent of the concentration and temperature.  The occupancy variation with
$x$ and $T$ is significantly larger for the other sites.  It was found that
increasing of $x_{Fe}$ in the alloy results in an increase of Fe
content mainly on the sites C and E.  For the latter, one can observe the highest
sensitivity to temperature ($P_E$ decreases with temperature) in comparison to
much slower variations revealed for B and C site occupancies (that increase with
temperature).

It should be noted that the entropy contribution that predominantly effects
the sublattice occupancies is the configuration entropy term.  Neglecting $S_{mag}$ in
the calculation of $P_{i}$, depending on $x$ and $T$ as shown in Fig.  4, leads to
very small changes.  It becomes clear when accounting for a
similarity of all individual curves presented in Fig.  2.  There, $S_{mag}$ curves vary mainly
with the total concentration of Fe/V atoms, and are very similar for all sublattices.
This means that this form of the entropy does not substantially affect the site preferences.

The solid lines (Fig.  4) indicate the $P_{i}$-values calculated for different
compositions of the $\sigma$-Fe$_{100-x}$V$_x$ with $x$ ranging from 33 to 60
(full range of $\sigma$-FeV occurrence).  Vertical bars - marked in the plots - indicate
values of $P_{i}$ as determined at RT, but for the samples transformed to the
$\sigma$-phase at 1300 K (that is why the bars are marked at this temperature
in the plot). As the figure shows, the site occupancies calculated by the minimization
of the free energy fairly agree with the values obtained from the neutron diffraction
experiments.  The agreement between theory and experiment is less satisfying
for the site E, since the calculated $P_{i}$-curves span over a wider range than the
the width of the corresponding bar.

The discrepancy observed between the calculated and experimental
$P_{i}$-values is likely due to several factors. One of them is the simplified model used for the entropy
calculations viz. without taking into account the phonon and the electron entropy terms.
However, a possible influence of these terms on final results seems to be rather small as
previously found from the formation energy analysis of the $\sigma$-phase in
the Fe-Cr alloy system \cite{Cieslak12a}.  Unfortunately, we are not aware of any
literature data reporting vibrational properties of the $\sigma-$FeV that would
allow to estimate the $S_{ph}$ contribution.  On the other hand, bearing in mind a similarity
of atomic and lattice parameters for Fe-V and Fe-Cr $\sigma$-phases, it seems
reasonable to assume that the $S_{ph}$ contribution to the total entropy is also small in the present case.
The effect of the $S_{el}$ term was not taken into account mostly due to the fact
that our discussion was focused on temperatures $\sim 1300$ K that are far above the
Debye temperature of the investigated samples ($\sim400-600$ K \cite{Cieslak10c}).

Another factor that could be responsible for the discrepancy between the calculated
and the experimental $P_{i}$ - values is the assumed form (polynomials) of the functions
for the $\Delta E$ and $S_{mag}$ (hence, in consequence that of
$\Delta F$). Better fits were obtained by increasing
the degree of the used polynomials, but in that case the discrepancy was not improved.

\section{Summary}

The formation energy, $\Delta E$, magnetic, $S_{magn}$, and configuration, $S_{conf}$, entropy contributions
were calculated and analyzed in the full range of the FeV-$\sigma$-phase
existence using the electronic band structure calculations by means of the KKR method.
It was found that $\Delta E$-values strongly depend on the Fe concentration, and their
variation observed for different site
occupancies is characteristic of a given lattice site.  The changes also
strongly depend on the number of Fe atoms on the sites.  Similar, increasing dependence on the
sample's composition exhibits the $S_{magn}$ term, but for a given composition it weakly depends on
the lattice site. On the other hand, the composition dependence of the $S_{conf}$ term was found to be weak,
although for a given composition, $S_{conf}$, shows well-defined dependence on the lattice site, and for each site,
a rather strong dependence on the number of Fe atoms occupying the site.
The sublattice occupancies were determined for various compositions and
temperatures by minimizing the free energy.  The occupancies of A and
D sites were found to be almost independent of $T$ and $x$, while the occupancies of the sites B, C
and D showed a higher sensitivity to $T$ and $x$. The KKR calculations combined
with the analysis of the entropy contributions clearly showed that
Fe atoms preferably occupy A and D sites in the FeV-$\sigma$ phase. This agrees well
with the neutron diffraction measurements, and it confirms the general observation for $\sigma$
in binary alloy systems that the smaller atoms (in this case Fe) tend to mostly reside on the sites
having the shortest distance to the nearest-neighbors (A and D), while the bigger atoms (here V) prefer to occupy the sites
having more fee space to accommodate them (B, C, E).

\begin{acknowledgments}
This work, supported by the European Communities under the contract of
Association between EURATOM and IPPLM, was carried out within the framework
of the European Fusion Development Agreement. It was also supported by the
Ministry of Science and Higher Education, Warsaw (grant No. N N202 228837).
We also acknowledge financial support from the Accelerated Metallurgy Project,
which is co-funded by the European Commission in the 7th Framework Programme
(contract NMP4-LA-2011-263206),
by the European Space Agency and by the individual partner organizations.

\end{acknowledgments}

\appendix

\section{Nomenclature}

There are several possibilities for description of sublattice occupancy in the FeV $\sigma$-phase.

One can directly indicate $n_i$ ($i=A$ through $E$), the number of Fe atoms occupying each site.
Using that, ${\sum n_i} = n_{Fe}$ gives the total number of Fe atoms in the unit cell
and ${{1\over 30} \sum n_i}$ gives the total concentration of iron.

Sublattice occupancy by Fe atoms can be easily calculated here as $x_i = {n_i/n_{Fe}}$
(in other words $x_i$ stands for the probability that randomly chosen Fe-atom belongs to the sublattice~$i$,
$\sum x_i = 1$).

On the other hand, $P_i = {n_i\over N_i}$ ($N_i$ being the multiplicity of the $i-$th sublattice)
denotes Fe-concentration on $i$-th sublattice (the probability, that randomly chosen atom on $i$-th
sublattice is iron).

\section{Polynomial coefficients}
From the fitting procedure $\Delta E$ (in [eV/atom]), and $S_{mag}$ (in [eV/K/atom$\times 10^{-8}$]) were determined,
as follows:

\begin{eqnarray}
\Delta E =
 0.17446
-0.03178n_1
+0.01269n_2                        \nonumber \\
-0.03404n_3
+0.00768n_4
-0.00689n_5                        \nonumber \\
+0.00258n_5^2                      \nonumber \\
+0.00226n_3(n_1+n_2+n_4+n_5)      \nonumber \\
-0.00160n_5(n_1+n_2+n_4+n_5)
\end{eqnarray}

\begin{eqnarray}
S_{mag} =
-1525.45
-199.617n_1
+247.216n_2          \nonumber \\
+195.366n_3
+624.922n_4
+137.535n_5          \nonumber \\
+62.3663n_5^2
-46.5917n_4^2
-11.4958n_3^2        \nonumber \\
-56.8196n_2^2
+117.045n_1^2        \nonumber \\
+287.426n_1n_2
+35.9117n_1n_3      \nonumber \\
-72.3616n_1n_4
+0.05473n_1n_5      \nonumber \\
+41.4487n_2n_3
-3.62722n_2n_4      \nonumber \\
-63.1291n_2n_5
+95.9298n_3n_4      \nonumber \\
-7.77130n_3n_5
-38.1694n_4n_5
\end{eqnarray}

\end{document}